\documentclass[aps,pre,twocolumn,superscriptaddress]{revtex4}
\usepackage{epsfig}
\usepackage{amsfonts}
\usepackage[figuresright]{rotating}
\usepackage{amssymb}
\usepackage{amsmath}
\usepackage{subfigure}
\usepackage{multirow}
\usepackage{tabularx}
\usepackage{bm}
\usepackage{array}
\usepackage[usenames]{color}

\usepackage{graphicx}

\def \indreal {\mathbf{n}} % The discrete position index in real space
\def \indrecip {\mathbf{q}} % The discrete wavenumber index in reciprocal space

\def \ireal {n} % The 1d real index
\def \irecip {q} % The 1d reciprocal index
\def \latvec{\textbf{a}} % a lattice (Bravais) vector
\def \recipvec{\textbf{G}} % a reciprocal lattice vector
\def \basisvec{\textbf{b}} % a basis vector

\def \evec {\textbf{e}} % The vector of extensions
\def \uvec {\textbf{u}} % The vector of displacements
\def \ext {e} % A single extension
\def \disp {u} % a single displacement
\def \fvec {\textbf{f}} % The vector of forces
\def \rmat {\textbf{R}} % The rigidity matrix
\def \dmat {\textbf{D}} % The dynamical matrix

\def \dim {d} % The dimension of space
\def \dimlat {d} % dimension of lattice
\def \length{N} % The length of the system in the number of sites, an integer, which we just label N
\def \toppol{\mathbf{P}^T}
\def \nzero{N_0} % number of zero modes
\def \ncon{N_{\textrm{con}}}
\def \ndof{N_{\textrm{dof}}}
\def \nsite{N} % number of particles
\def \nss{N_{\textrm{ss}}} % number of self stresses 
\def \pos{\mathbf{n}}

\def \recip{\mathbf{q}}

\def \esource{\mathbf{e}_s}
\def \fsource{\mathbf{f}_s}

\def \pos{\mathbf{r}} % A position

\begin{document}
\title{Directional mechanical response in the bulk of topological metamaterials}

\author{D. Zeb Rocklin,
\\
Laboratory of Atomic and Solid State Physics, Cornell Univ., Ithaca, NY 14853-2501, USA
\\
School of Physics, Georgia Institute of Technology, Atlanta, Georgia 30332, USA
}

%\vspace{10mm}
\date{\today}

\begin{abstract}
Mechanical metamaterials are those structures designed to convey force and motion in novel and desirable ways. Recently, Kane and Lubensky showed that lattices at the point of marginal mechanical stability (Maxwell lattices) possess a topological invariant that describes the distribution of floppy, zero-energy \emph{edge} modes. Here, we show that applying force at a point in the \emph{bulk} of these lattices generates a directional mechanical response, in which stress or strain is induced only on one side of the force.  This provides both a bulk metric for mechanical  polarization and a design principle to convey stresses and strains towards or away from parts of the structure. We also characterize the effects of removing bonds on the material's structure and floppy modes, establishing a relationship between edge modes and bulk response. 
% Relying only on the topological structure of linear response, similar phenomena may be expected in such systems as origami/kirigami sheets, acoustic systems and frustrated antiferromagnets.
%Relying only on the topological structure of linear response, our results apply equally to ball and spring networks, origami/kirigami sheets, frustrated antiferromagnets, and to electrical circuits.
\end{abstract}

\maketitle

\iffalse

Wishlist: 

redo contour figure from index i to index j

redo lush figure so bonds are light gray, not black

Include cartoon of rotor system

\fi

\section{Introduction}

Mechanical metamaterials have engineered structures that imbue them with novel mechanical properties. Such structures can be auxetic (negative Poisson's ratio)~\cite{greaves2011poisson}, pentamode (fluid-like vanishing of most elastic moduli)~\cite{milton1995elasticity,kadic2012practicability} or even possess negative compressibilities~\cite{nicolaou2012mechanical}. The control of mechanical and acoustic response permits programmable elastic response~\cite{florijn2014programmable}, cloaking~\cite{milton2006cloaking,buckmann2014elasto} and soft robotics~\cite{kim2013soft}. Even beyond metamaterials, the mechanical response of natural and conventionally engineered materials is determined by their structures.

Of particular interest are Maxwell lattices~\cite{lubensky2015phonons} which exist at the isostatic point, possessing the minimal number of bonds necessary to support external force, just as do naturally occurring systems such as fiber networks~\cite{broedersz2011criticality} and jammed packings~\cite{liu1998nonlinear}. Each missing bond at the edge of a finite Maxwell lattice generates a floppy mode in which particles may move without stretching bonds. Kane and Lubensky~\cite{kane2014topological} revealed that Maxwell lattices possess a nontrivial \emph{topological polarization} vector $\toppol$ and that, remarkably, the bonds removed at one edge can generate zero modes on the far edge. Subsequently, it has been shown that this topological polarization couples to line-like~\cite{paulose2015selective} and point-like~\cite{paulose2015topological} defects to create new ways of distributing force and motion. Nonlinear modes have been examined~\cite{chen2014nonlinear,rocklin2015transformable} that can both couple to the topological polarization and change it. Systems which break time-reversal symmetry, such as gyroscopes~\cite{nash2015topological, wang2015topological}, acoustic resonators~\cite{yang2015topological} etc. have also been shown to have topological structure, in the form of a Chern-Simons number, that generates protected chiral edge modes.

Kane and Lubensky mapped Maxwell frames onto topological insulators, whose well-known bulk-boundary correspondence is associated with the topological edge modes. Their topological index may be expressed in terms of a bulk Green's function~\cite{essin2011bulk}, raising the question of
whether the bulk Green's functions of Maxwell frames reveal physical phenomena associated with their nontrivial topology.
Previous examination of the bulk elastic properties of Maxwell frames~\cite{calladine1978buckminster, guest2003determinacy, lubensky2015phonons} has identified a characteristic mode, the ``Guest mode'', which deforms the lattice uniformly without stretching any bonds. This mode is present for any Maxwell lattice, and is hence a signature of instability but not of polarization.

In the present work, we show that there is a topological response that exists alongside this long-wavelength elastic behavior. Local forces applied to a polarized lattice generate either stress (for generic forces) or strain (for forces applied only along bonds) on \emph{only one side} of the applied force. Indeed, this bulk mechanical response can account for missing bonds, accurately encapsulating the anomalous mechanical behavior at the edges of polarized lattices.
The topological response is in stark contrast to the elastic response function of fully stable elastic media, in which stresses and strains necessarily extend in all directions.

The rest of our paper is arranged as follows. Sec.~\ref{sec:response} describes the general response (Green's) functions of mechanical lattices. Sec.~\ref{sec:toppol} discusses the implications of topological polarization. Secs.~\ref{sec:onedchain},\ref{sec:kagome} consider 1D chains and 2D kagome lattices, respectively. Sec.~\ref{sec:edgemodes} associates zero modes with missing bonds in the bulk and edge. In Sec.~\ref{sec:discuss} we discuss the implications of our work.

\section{Mechanical response functions in Maxwell lattices}
\label{sec:response}

\begin{figure}
\subfigure{\includegraphics[width=0.49\textwidth]{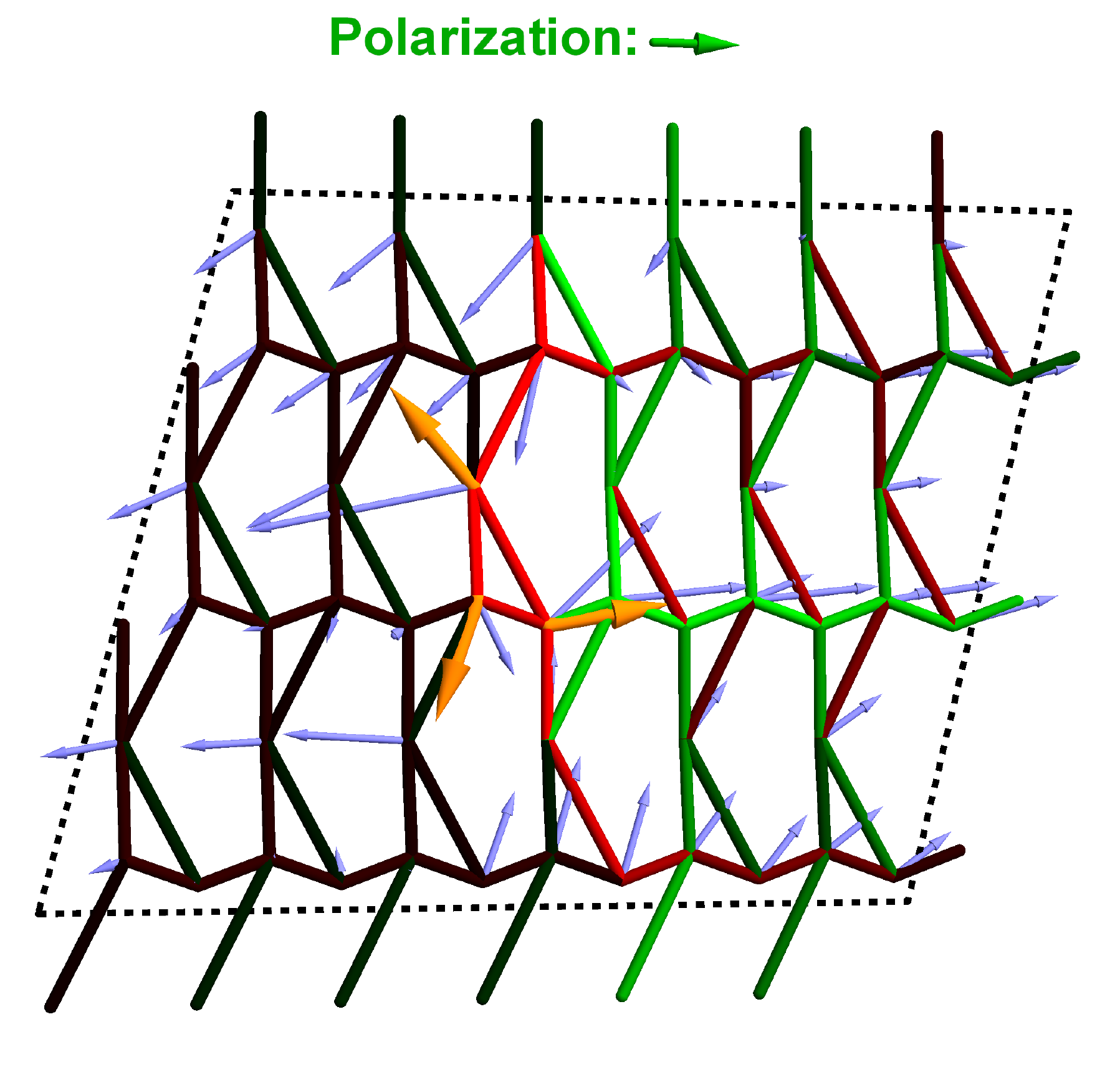}}
\caption{
A section in the bulk of a topological Maxwell system, the generalized kagome lattice. Externally-applied forces (orange arrows) shift sites along blue arrows in all directions. However, for the polarized lattice bond extensions (red) and compressions (green) are generated only in the direction of polarization.
}
\label{fig:lushfig}
\end{figure}

We consider a system of the form shown in Fig.~\ref{fig:lushfig}, consisting of a periodic lattice of particles connected by bonds in $\dim$ dimensions. For simplicity, we take masses and Hookean spring moduli to be unity. Restricting ourselves to small strains, we can linearly map the vector of site displacements $\uvec$ onto the vector of bond extensions $\evec$:

\begin{align}
\label{eq:extensions}
\evec = \rmat \uvec.
\end{align}

\noindent $\rmat$, which we refer to as the rigidity matrix (also ``compatibility'' or ``kinetic'' matrix), is determined by the sites' equilibrium positions and bond structure, as described in Appendix A. Because the tensions of our springs are proportionate to their extensions and generate force along the bond directions, we also have the vector of forces $\fvec$ given by

\begin{align}
\label{eq:forces}
\fvec = \rmat^T \evec.
\end{align}

\noindent Combining these relationships leads to the energy functional $E$ and to the relationship between forces and displacements, both in terms of the dynamical matrix $\dmat = \rmat^T \rmat$:

\begin{align}
\label{eq:dmat}
\fvec = \dmat \uvec, \\
E = \frac{1}{2} \uvec^T \dmat \uvec.
\end{align}

Linear zero modes, whose number we term $\nzero$,
 are sets of displacements which do not stretch any bonds and thus lie in the nullspace of $\rmat$. Sets of bond tensions or elongations which do not generate net forces similarly lie in the nullspace of $\rmat^T$; these ``self stresses'' number $\nss$. The relationships of Eqs.~(\ref{eq:extensions}, \ref{eq:forces}) then lead to the following index theorem, given in terms of the system's numbers of degrees  of freedom ($\ndof$) and constraints ($\ncon$)~\cite{calladine1978buckminster,lubensky2015phonons}:

\begin{align}
\label{eq:index}
\ndof - \ncon = \nzero - \nss.
\end{align}

\noindent In particular, for Maxwell lattices zero modes may be generated only in conjunction with self stresses or by removing bond constraints.

We now ask what the effects on the lattice will be of either applying forces or of \emph{swelling} certain bonds.
Swelling is any static or quasistatic process that changes the equilibrium length of a bond, including fixed length disorder, dynamic fluctuations, external forces, or mechanical actuation  (the concept of altering equilibrium lengths is implicit in self stresses, wherein tension is induced in bonds without changing their lengths).
Although such analysis may also be applied to disordered systems, we now restrict ourselves to periodic structures described by $\dimlat$ lattice vectors in a $\dim$-dimensional space (for some systems, such as origami, these two numbers are not equal, though this does not affect the analysis).
The response in a lattice, the 
\emph{lattice Green's function}, is well-developed (see, e.g.,~\cite{katsura1971lattice,economou1984green}). Unlike previous studies of response in spring networks, the present work follows the concept of Kane and Lubensky in taking the ``square root'' of the Green's function and examining the effect of bonds on sites and vice-versa, thereby revealing topologically-nontrivial behavior.

The position of crystal cells in the lattice is indexed by a discrete vectorial index $\indreal$ and, in reciprocal space, a wavenumber $\indrecip$. Due to our local interactions, reciprocal-space succeeds in diagonalizing the rigidity map. Taking our wavenumber continuous, as is valid 
even over short distances
in large systems, leads to maps of the form

\begin{align}
\label{eq:rmatq}
\evec = \rmat(\indrecip) \uvec,
\end{align}

\noindent etc. Now, $\uvec$ refers to the mode within a single cell, such that $\uvec_\indreal = \exp(i \indrecip \cdot \indreal)\uvec$.
Inverting the above relationship provides us with the response of the lattice to the swelling of a single bond, $\evec_\indreal = \evec_s \delta_{\indreal,\indreal_s}$:

\begin{align}
\label{eq:inverse}
\uvec_\indreal = \frac{1}{\left(2\pi\right)^\dim}\int_{-\infty}^{\infty} d^\dim \indrecip \,
 e^{i \indrecip \cdot \left(\indreal-\indreal_s\right)}\rmat^{-1,g}(\indrecip) \esource.
\end{align}

\noindent 
We may similarly invert Eqs.~(\ref{eq:forces},\ref{eq:dmat}) to
 relate $\uvec$ to $\fvec$ via $\dmat$ and $\evec$ to $\fvec$ via $\rmat^T$. Here, $\rmat^{-1,g}$ denotes the Moore-Penrose generalized pseudo-inverse, which may be readily generated from singular-value decomposition even for non-square matrices~\cite{dresden1920,ben2003generalized} corresponding to non-Maxwell systems. This set of site displacements satisfies the linear map of Eq.~(\ref{eq:extensions}) when possible and more generally is the best possible solution in the sense of minimizing the magnitude of the additional extensions, $\rmat \uvec - \evec$, and hence the energy of the lattice.

This failure to be fully invertible is not, in general, a minor issue. For a general bond network, these additional extensions are the tensions resultant from applied forces. However, general dynamical matrices and rigidity matrices of Maxwell lattices are square matrices which possess only those zero modes that arise from finely-tuned geometric features or from global symmetries. Lattices have $\dim$ such translational symmetries (our boundary conditions prevent global rotations) and hence via the index theorem [Eq.~(\ref{eq:index})] there are $\dim$ self stresses. There are a number of ways to address this issue:

\noindent 1) We consider sets of modes that don't include translations, so that the rigidity matrix is exactly invertible, as in Sec.~\ref{sec:onedchain}.

\noindent 2) We swell combinations of $\dim+1$ bonds chosen to avoid the $\dim$-dimensional nullspace of $\rmat$, as in Sec.~\ref{sec:kagome}.

\noindent 3) We consider applied forces, rather than swollen bonds, so that inversion of $\dmat(\indrecip)$ is ensured provided that the  total force on the structure is zero (i.e., force dipoles), as in Sec.~\ref{sec:kagome}.

\noindent 4) We consider signals occurring only at a particular wavenumber $\indrecip \ne 0$, so that $\rmat(\indrecip)$ is invertible, as in Sec.~\ref{sec:edgemodes}.

Upon applying one of these techniques, the Maxwell lattice, unlike fully rigid systems with more bonds, has essentially a one-to-one correspondence between strained shape and the pattern of stress throughout its structure. This means in particular that a desired configuration, given to linear order by $\uvec_\textrm{des}$, may be achieved in a simple, controlled fashion by using mechanical actuation to extend and contract bond lengths to $\evec = \rmat \, \uvec_\textrm{des}$.

\section{Topological polarization and directional response}
\label{sec:toppol}

As discussed in Appendix B, a Maxwell lattice always possesses some number of zero modes set by the number of bonds in a unit cell: they simply aren't always normalizable. For example, for the 1D chain considered in the next section, the sole zero mode occurs at complex wavenumber such that $\exp(i \irecip) = b/a$ and hence grows or shrinks as one moves across the lattice. These modes violate periodicity, but for a finite lattice
with open boundaries
 a zero mode that, e.g., is growing as one moves to the right is present on the right edge. Kane and Lubensky~\cite{kane2014topological} discovered the remarkable result that these edge modes are not determined solely by the local coordination number (i.e., by how many bonds were removed) but also by the topological polarization,

\begin{align}
\label{eq:toppol}
\toppol_j = -\frac{1}{2 \pi i} \int_{\indrecip \rightarrow \indrecip + \recipvec_j}
\frac{\partial}{\partial \irecip_j} \log \textrm{det} \left[\rmat(\indrecip)\right],
\end{align}

\noindent with $\recipvec_j$ the $j^\textrm{th}$ vector of the reciprocal lattice. This beautiful relationship means that, once gauge and boundary conditions are fixed~\cite{kane2014topological}, the winding of the bulk finite-energy modes determines the placement of the edge zero modes.
Generically, $\toppol$ in fact may differ at different wavenumbers~\cite{rocklin2016mechanical},
but we do not address this case in the present work. 

We focus on the particular case in which \emph{every} zero mode of a lattice shrinks (or grows) along a lattice direction $\latvec_j$. We call this special case of polarization ``fully polarized''. These correspond to zero modes with  $\textrm{Im}(\irecip_j)>(<)0$, as shown in Fig.~\ref{fig:contour}.
Thus, by considering the poles of $\rmat^{-1}(\indrecip)$ as shown in Appendix B, we find that 
 \emph{when a bond is swollen in a fully polarized lattice 
displacements of the polarized lattice will be localized on only the unpolarized side}. Repeating the analysis for applied forces and noting that for every zero mode of $\rmat(\indrecip)$ there is a self stress of $\rmat^T(-\indrecip)$, we see that similarly, \emph{when a force is applied in a fully polarized lattice bonds will be stressed on only the polarized side}, as shown in Fig.~\ref{fig:lushfig}.

\begin{figure}
\subfigure{\includegraphics[width=0.49\textwidth]{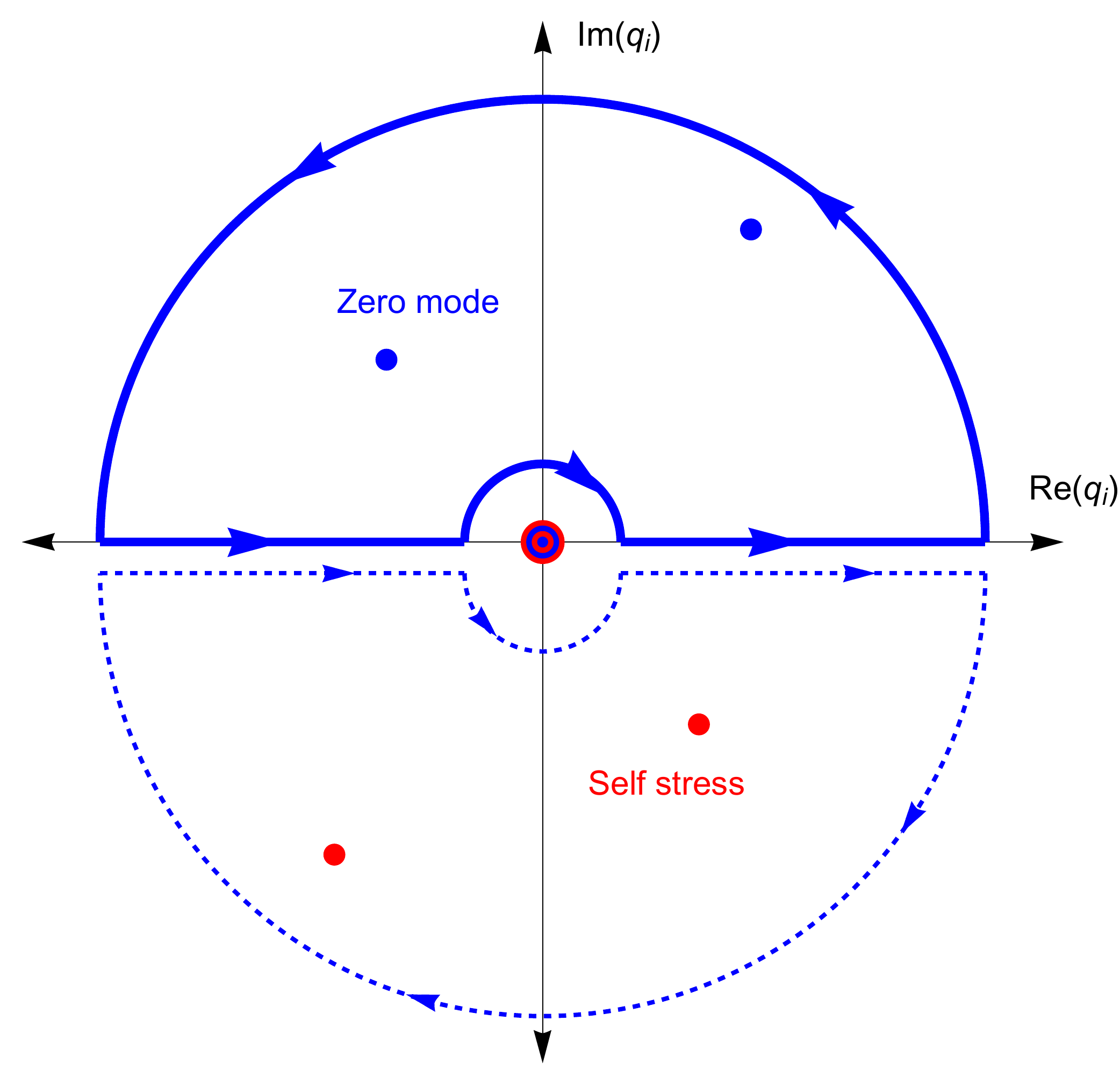}}
\caption{
The nullspace of the dynamical matrix possesses states of mechanical equilibrium, zero modes and self stresses, at complex wavenumbers. These modes appear as poles of $\dmat^{-1}(\indrecip)$ in the complex plane that result in contributions to the response functions. The response then combines the long-wavelength elastic response from the states at $\irecip_i = 0$ and the states at finite $\irecip_i$ enclosed by the contour. The different contours (solid and dashed) are valid for obtaining the response on opposite sides of the source and hence for topologically polarized systems wherein all zero modes are enclosed by one contour the response vanishes on one side of the source.
}
\label{fig:contour}
\end{figure}

\section{One-dimensional topological chains}
\label{sec:onedchain}

\begin{figure}
\subfigure{\includegraphics[width=0.49\textwidth]{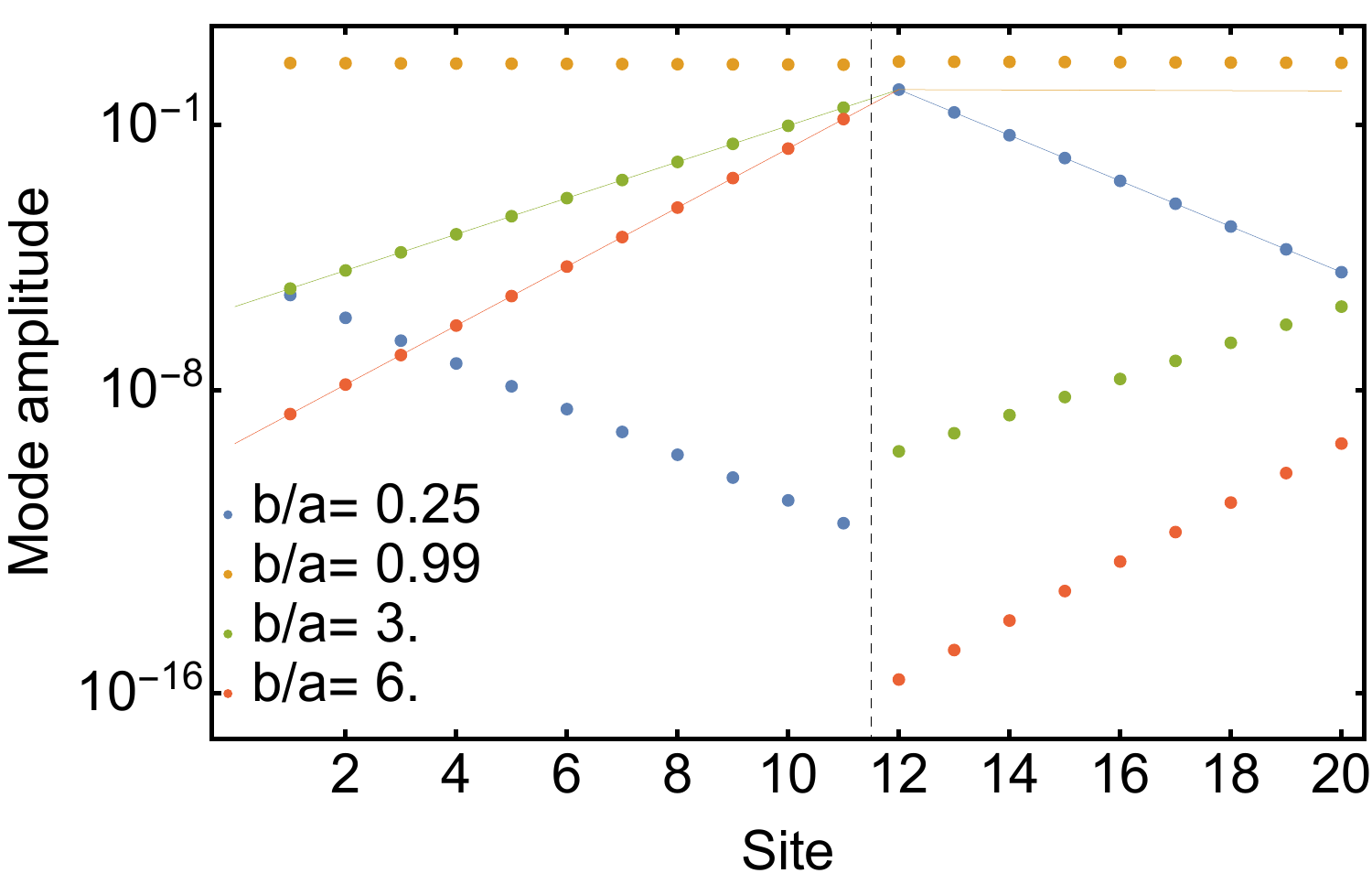}}
\caption{
The deformation modes of a topological 1D chain of 20 sites with a bond swollen between the eleventh and twelfth sites. The modes are exponentially localized on a side determined by the topological polarization. Direct calculation (points) and contour integration (lines) are in good agreement except when the localization length becomes comparable to system size (upper, orange points).}
\label{fig:chain}
\end{figure}

We now illustrate the general result of the preceding section by considering the simplest topological response function. For the 1D topological chain of rotors used by Kane and Lubensky~\cite{kane2014topological}, the extension of a bond is related to the displacement of two rotor heads via

\begin{align}
\label{eq:onedchain}
e_\ireal = a u_\ireal - b u_{\ireal-1}.
\end{align}

\noindent Beyond the rotor system, this represents
the most general uniform one-dimensional mapping from a mode $\uvec$ to a set of energetically costly deformations $\evec$, also describing for example an origami chain~\cite{chen2016topological}.
 We now consider the effect of swelling a single bond, i.e. requiring $\ext_{\ireal'} = \ext_s \delta_{\ireal',\ireal_s}$. For such a simple system, we may for the finite case explicitly invert Eq.~(\ref{eq:onedchain}) and for the infinite case perform a contour integration, the latter resulting in

\begin{align}
\label{eq:onedmode}
\disp(\ireal)=\frac{1}{a}\left(\frac{b}{a}\right)^{n-n_s} \textrm{sign}(n-n_s)\Theta\left[\left(n-n_s\right)\left(a-b\right)\right].
\end{align}

\noindent We see that the displacements are exponentially localized on only \emph{one} side of the swollen bond. That side is determined by the location of the zero mode $\exp(i \irecip) = b/a$ in the complex plane, which also determines the topological polarization. The resultant mode is shown in Fig.~\ref{fig:chain}, in good agreement with the results obtained from direct calculation of systems with periodic boundary conditions and only a handful of sites $\nsite=20$. Indeed, this form of the mode holds when $(b/a)^\nsite \ll 1$ or $\gg 1$, when the mode vanishes before the boundary. For the case $a=b$, as in a simple chain of balls and springs, the swelling of a single spring results in displacements that do not diminish with distance. In all other cases, the bulk mode of Eq.~(\ref{eq:onedmode}) is precisely (up to rescaling) the zero edge mode obtained by Kane and Lubensky~\cite{kane2014topological}: allowing a single bond in a periodic system to swell or contract freely permits a mode wherein the site to the right (left) experiences a missing bond to its left (right), precisely the condition experienced by a system with open boundaries. Just as the topological polarization allows a single deformation mode to grow in one direction towards the edge, it allows one to grow towards the swollen bond.

\section{Generalized kagome lattice}
\label{sec:kagome}

We now consider a two-dimensional, translationally invariant Maxwell frame, the generalized kagome lattice. The crystal cell consists of three sites connected via six bonds to each other and to three neighboring cells in such a way that no bonds are parallel to one another, as shown in Fig.~\ref{fig:lushfig}. Unlike the 1D chain of Sec.~\ref{sec:onedchain}, the kagome lattice possesses the general features of a Maxwell lattice: a non-scalar rigidity matrix, multiple lattice directions and translational zero modes and attendant states of self stress. In considering the response to the swelling of a bond, the first two features preclude a simple analytic expression as obtained in Eq.~(\ref{eq:onedchain}), but are still readily addressed numerically. The self stresses, however, require additional consideration. 

\begin{figure*}
\subfigure{\includegraphics[width=0.98\textwidth]{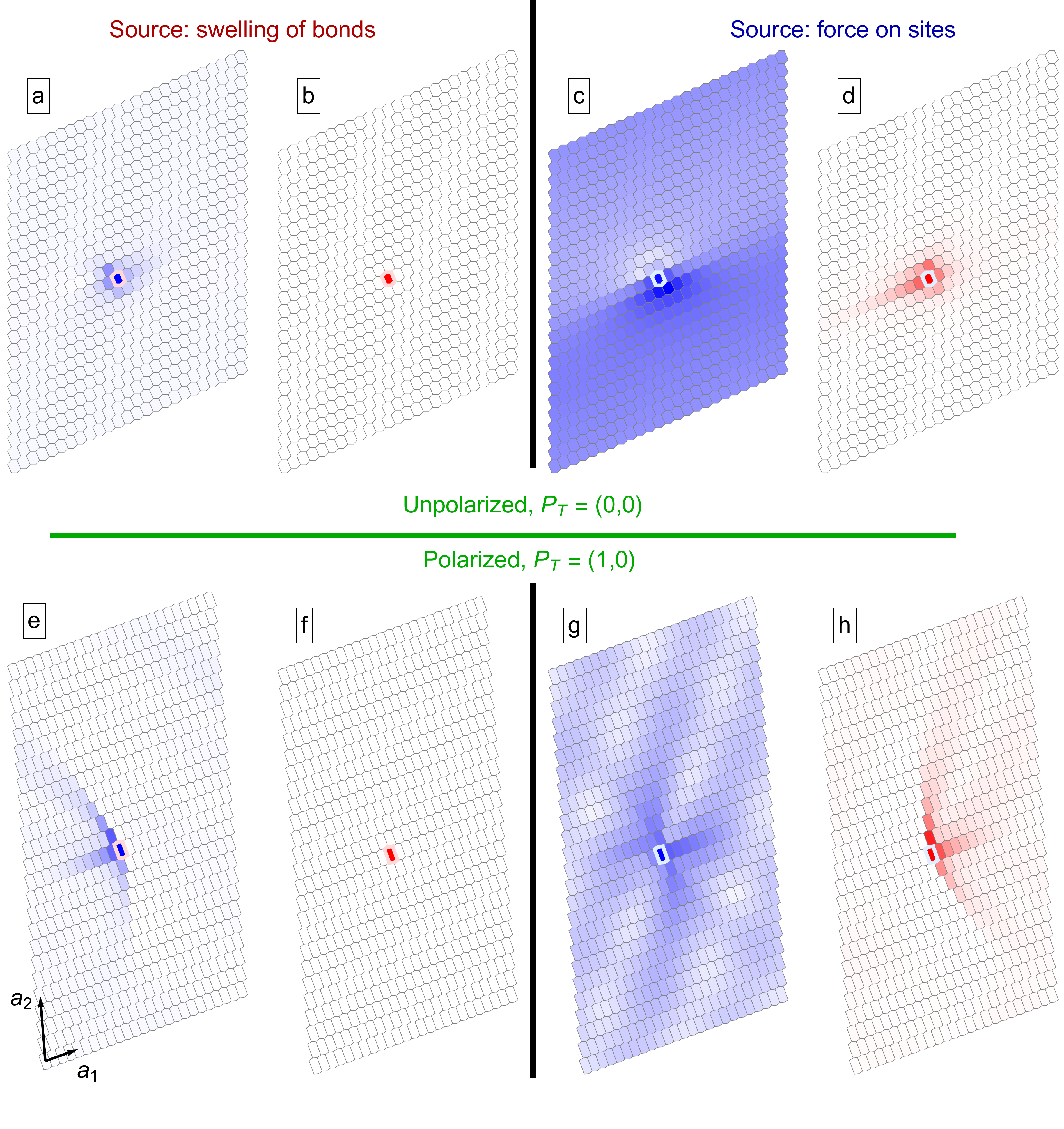}}
\caption{
The response of unpolarized (top row) and polarized (bottom row) generalized kagome lattices to swelling bonds (left two columns) and applied forces (right two columns). Hexagons are unit cells with bonds and sites not shown; blue color indicates intensity of displacements while red color indicates intensity of bond extensions/compressions. The central red-highlighted cells have bond swellings externally induced, while blue-highlighted cells have forces applied.
Induced swellings are chosen to avoid self stresses as described in the main text so that when bonds are swollen in (b), (f) no tension is induced in other cells. The principal effect of topological polarization is visible in (e), where sites move on only one side of swollen bonds and (h), where bonds stretch only on one side of applied forces. In contrast, the response in (f), (g) remains symmetrical.
}
\label{fig:mainfig}
\end{figure*}

%When we apply forces, as in Fig.~\ref{fig:mainfig}(c),(d),(g),(h), the issue does not arise.
 When we swell a single bond, the pseudoinversion of Eq~(\ref{eq:inverse}) generates a spatially extended projection into the two global self stresses. Instead, to exemplify the fully-polarized result of Fig.~\ref{fig:mainfig}(e),(f), we swell three bonds in a combination selected to cancel out the contribution to the two self-stresses. The result is that, as shown in Fig.~\ref{fig:mainfig}(b),(f), none of the remaining bonds in the lattice are extended. Hence, we see the effect of full topological polarization described in the previous section: when bonds are extended sites shift position only to the left [Fig.~\ref{fig:mainfig}(e)] whereas when forces are applied to sites bonds are extended and compressed only to the right [Fig.~\ref{fig:lushfig} and Fig.~\ref{fig:mainfig}(h)].

The topological polarization indicates that the zero mode extends only to cells for which $\ireal_1 < \ireal_1^s$. However, in terms of the lattice directions $\latvec_1,\latvec_2$, a triangle in the kagome lattice couples also along $\latvec_2 - \latvec_1$, permitting us to calculate the topological polarization in a new crystal basis. The result separately indicates that the zero mode is along $\ireal_2-\ireal_1<\ireal_2^s-\ireal_1^s$ and so one can see in Fig.~\ref{fig:mainfig}(e),(h) that for the kagome at least the topological polarizations may be used to restrict a response not to one half of the plane but to one third. Hence, topological polarization in the original basis does not fully determine the bulk response of the kagome lattice:
the analogous phenomenon in terms of the edge modes would be the observation that the number of edge modes along a cut parallel to $\latvec_2 - \latvec_1$ is not determined solely by the topological polarization along the two original lattice directions.

For the 1D chain, we saw that the directionality, defined as the ratio of the mode intensity immediately on the polarized side of the source to that on the unpolarized side, was exponential in system size. Despite the complication of long-wavelength contributions in the transverse direction, the modes of 
two-dimensional lattices such as that of 
Fig.~\ref{fig:mainfig} (e)
are roughly exponentially localized as well. A typical fully polarized $40\times40$ lattice with periodic boundary conditions has a directionality on the order of one million.

Finally, we note that our bond swelling mode bears a striking resemblance to the soft modes observed in the same class of lattices when subject to a defect with a Burger's vector aligned in a certain way with the topological polarization~\cite{paulose2015topological}. Insofar as neither the swelling modes nor the defect modes stretch any bonds beyond the source, it may be that these are the same and the unique set of modes which do not stretch any bonds beyond a local disturbance.

\section{Bond modes and edge modes}
\label{sec:edgemodes}

As noted in Sec.~\ref{sec:onedchain}, the mode associated with swelling a bond in the periodic system resembles the mode on an edge of a chain with open boundary conditions,  which in fact reflects a more general phenomenon. 
We have obtained modes which swell certain designated bonds and no others; these must necessarily be the zero modes of a lattice in which the designated bonds are removed rather than swollen.
In the 1D chain, this simply means that removing the bond between the $\nsite^\textrm{th}$ site and the first site generates an open system with a zero mode on the left or right edge, and this is the same as the mode associated with the swelling of that same bond, which occurs wholly on either its right or left. Associating this displacement mode with a particular swelling or missing bond, we refer to it as a ``bond mode''.

In translationally-invariant systems, the picture is complicated by the zero modes and self stresses. Our method of choosing a combination of a few bonds at a time to avoid the self stresses does not generate a convenient basis. However, we may consider a bond vector $\evec$ at a particular wavenumber $\indrecip \ne 0$. Then, for general lattices (i.e., ones without extra self stresses resulting from redundant bonds) $\rmat(\indrecip)$ is invertible and we may define a particular bond mode $\uvec = \rmat^{-1}(\indrecip)\evec$. Similarly, we may define ``force modes'', sets of bond elongations resulting from forces on certain sites: $\evec = \rmat^{T}(\indrecip)^{-1}\fvec$.

\begin{figure}
\subfigure{\includegraphics[width=0.49\textwidth]{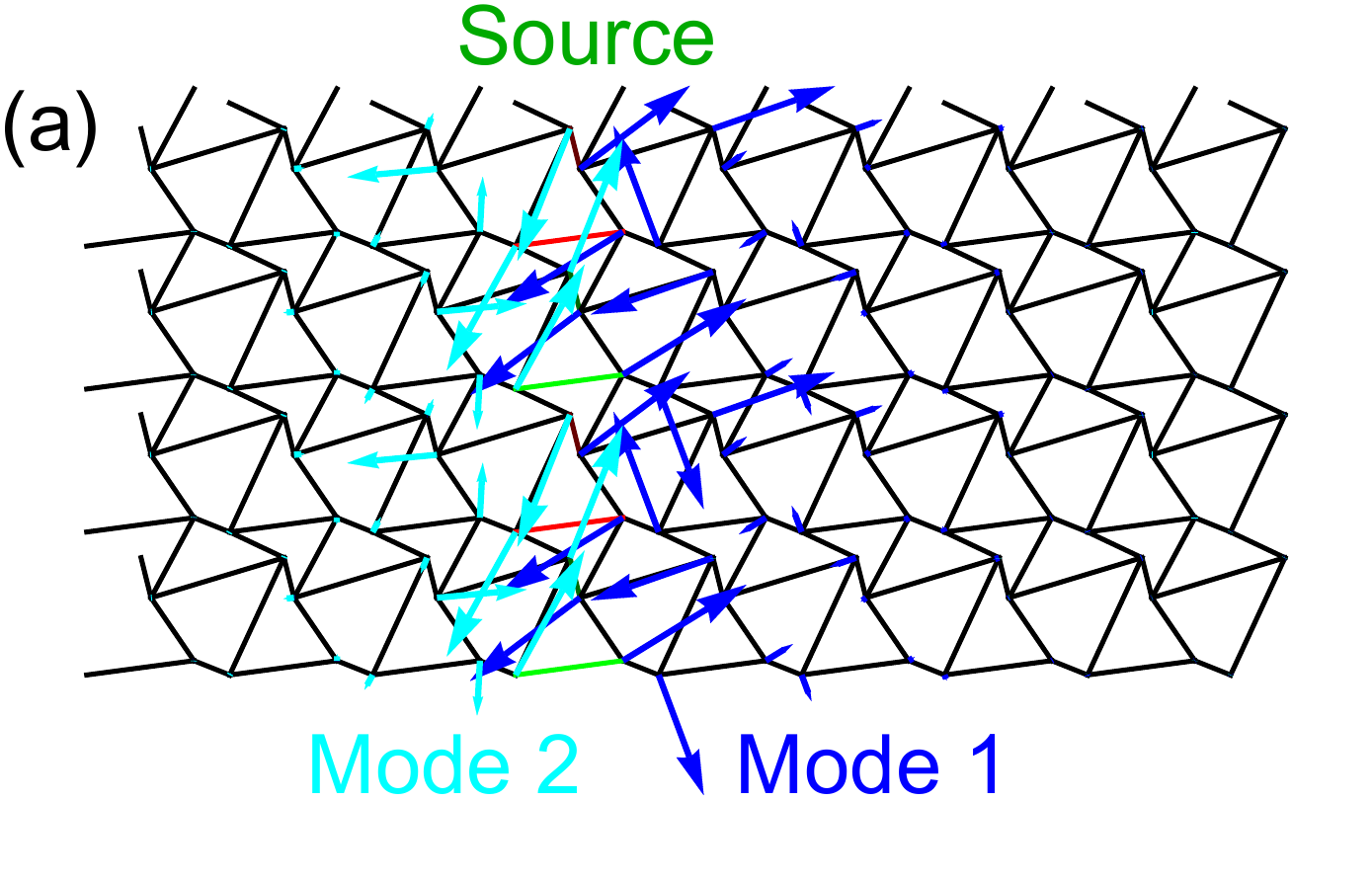}}
\quad
\subfigure{\includegraphics[width=0.49\textwidth]{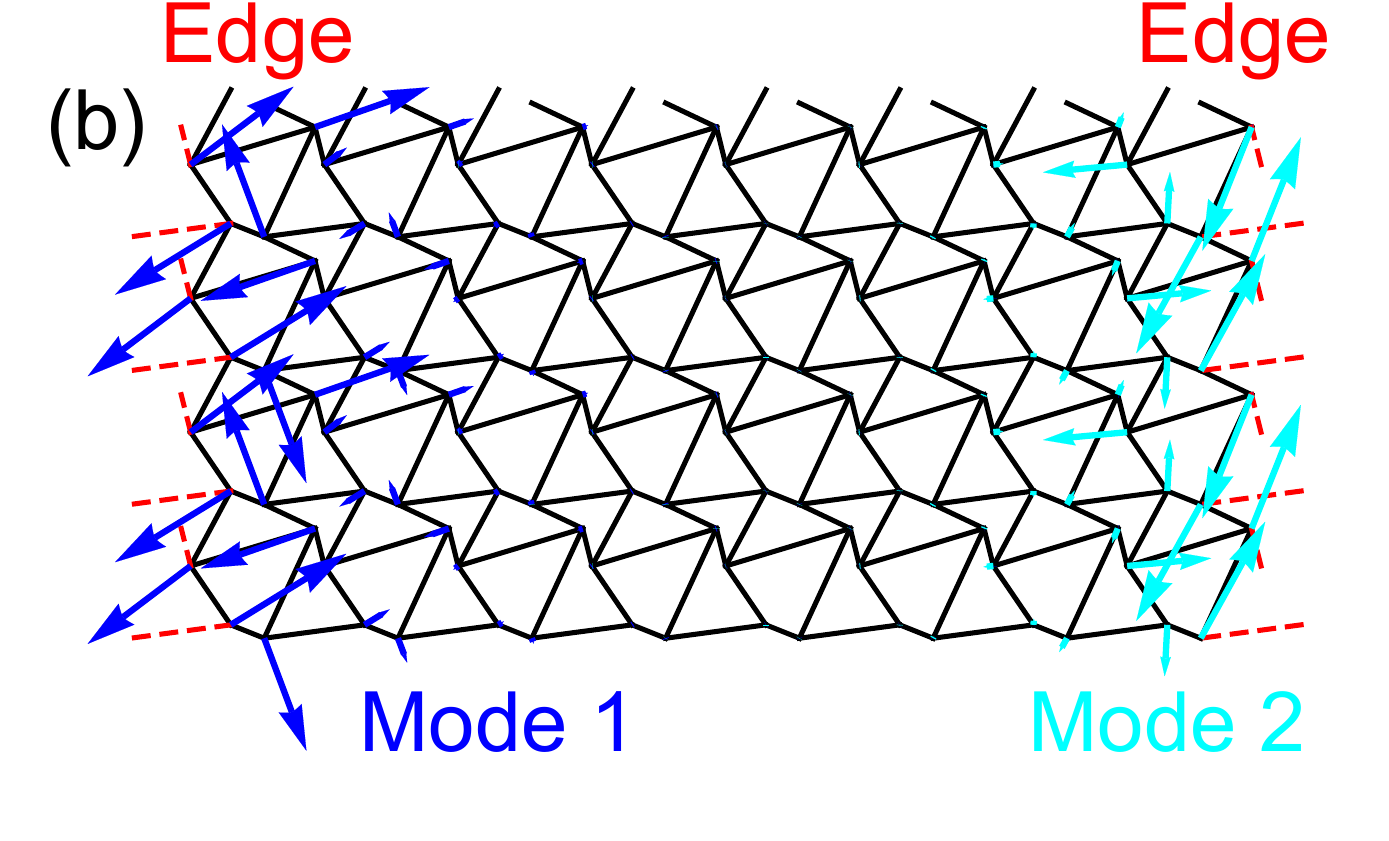}}
\caption{
(a) A deformed square lattice, unpolarized, with periodic boundary conditions subject to a column of swollen bonds with periodicity $\irecip_2 = \pi$ leads to a mode (blue arrows, right) that does not stretch additional bonds. A different swelling of the same column of bonds generates a second mode (cyan arrows, left). (b) A lattice with \emph{missing} rather than swollen bonds at its left and right edges has the \emph{same} $\irecip_2 = \pi$ zero modes as result from the swelling bonds, including direction and rate of decay.
}
\label{fig:edge}
\end{figure}

From the set of bond modes, we may now make a connection to the edge modes first associated with the topological winding number. Consider two sets of bond elongations, $\evec(\indreal) = \evec^{1,2}_s \delta_{\ireal_1,\ireal^s_1} \exp(i \irecip_2 \ireal_2)$, where the two bonds are those cut by a line drawn along the second lattice direction, as shown in Fig.~\ref{fig:edge} (a). Our response function of Eq.~(\ref{eq:inverse}) tells us that for $\recip \ne 0$ the modes resulting from these sets of bond swellings are exponentially localized either to the left or to the right of $\ireal_1^s$, with the number on either side determined by the topological polarization.
As in the 1D chain, these are also the modes on the right and left edges, respectively, of a large system with open boundaries. That is, we may use our bulk topological response function, which sees zero modes shifting from one side of a source to the other, to recapitulate the modes resultant from missing edge bonds that were first associated with the topological winding number, those shown in Fig.~\ref{fig:edge} (b).
This allows us to associate each edge mode to one of the missing bonds in the crystal cell (as in the 1D chain), or more generally with a linear combination of the same, as in Fig.~\ref{fig:edge} (a).
% This has the added utility of not only identifying which edge each zero mode lies on but of associating it to a particular missing bond at that edge.

The directional response of the bulk visible in Fig.~\ref{fig:mainfig}(e),(g) also explicates the edge modes of the lattices. In (e) the  mode that stretches only to the left of a swollen bond also is a zero mode on the right edge of a finite lattice, and the lack of any such modes stretching to the right reflects the lack of zero modes on the left edge of such a lattice. Similarly, the swollen bonds only on the right of an applied force in (g) reveal that a force on the left edge will stretch bonds and cost energy, while one on the right will displace sites without energetic cost. This distinction has been shown~\cite{rocklin2015transformable} to increase the edge stiffness exponentially.

\section{Discussion}
\label{sec:discuss}

We have shown that the physical implications of the topological winding number identified by Kane and Lubensky for Maxwell lattices extend beyond the edge modes to the bulk directional response of the lattice to swelling bonds. Similarly, analogously to the edge self stresses, the bulk response to applied forces in fully polarized lattices is for tension to be generated in bonds on only one side of the force. This counter-intuitive result is possible because the site displacements and bond extensions live in separate, though closely linked, vector spaces [or, in Kane and Lubensky's formulation, in distinct subspaces of the combined vector space $(\uvec,\evec)$].
% Although it is known generally that crystals that lack inversion symmetry have response functions which lack inversion symmetry **, the degree of directionality for the Maxwell lattices, exponentially strong for any systems where the decay length is below the system size, is remarkable.

We have considered the linear, zero-frequency local modes associated with topologically polarized Maxwell lattices. There are, however, a rich variety of other other sets of zero modes present in Maxwell lattices. Weyl lattices possess isolated pairs of zero modes at $\indrecip = \pm \indrecip_w$~\cite{rocklin2016mechanical} while other lattices possess curved lines of zero modes in the Brillouin Zone~\cite{sun2012surface,rocklin2015transformable,chen2016topological,power2016mechanical}.
%Origami is a two-dimensional lattice embedded in three dimensions with a ``realness'' condition~\cite{chen2016topological,power2016mechanical}. 
Generalized pyrochlore lattices, 3D isostatic systems, possess Weyl lines~\cite{stenull2016topological}, curved lines of zero modes. General Maxwell lattices transition from ``shear-dominant'' states with two lattice directions with vanishing speed of sound to ``dilation-dominant'' states without these soft directions~\cite{rocklin2015transformable}. Any of these may display novel mechanical response apparent in their linear Green's functions. In particular, because topological polarization occurs at finite complex wavenumber it does not appear to influence long-wavelength elasticity in a direct fashion; in contrast, many of these other systems possess a number of zero modes at small wavenumbers. Beyond the Maxwell frames studied here, there are a host of other topological mechanical systems, characterized by varying dimensionality and symmetry~\cite{susstrunk2016classification}. An examination of some or all of their response functions may yield significant results.

Indeed, the ``Maxwell'' condition is not limited to mechanical systems like ball-and-spring lattices or origami. It is, rather, a statement that the two coupled vector spaces, displacements $\uvec$ and extensions $\evec$, are of the same size. Any such systems with periodicity and without inversion symmetry may possess similarly directional response functions. 
Similar topological physics to the mechanical lattices has been considered in systems of magnons~\cite{lawler2016supersymmetry,owerre2016magnon,owerre2016topological}, electrical circuits~\cite{ningyuan2015time,albert2015topological} and photonics~\cite{lu2014topological}.

Our formulation linking topological polarization to the directionality of the displacements resulting from applied stresses also suggests a definition of mechanical polarization extending beyond regular, isostatic lattices. 
It has already been shown that even in regular lattices topological polarization can become wavenumber-dependent~\cite{rocklin2016mechanical} and can acquire, e.g., half-integer values in the original formulation as the crystalline symmetry of the lattice is reduced~\cite{rocklin2015transformable}.
Isostatic systems without any lattice symmetry should have significant response to bond swelling, and one could examine jammed packings~
\cite{liu1998nonlinear}, rigidity percolation models representing fibrous systems~\cite{jacobs1995generic}, disordered versions of triangulated origami~\cite{chen2016topological}, etc. Perhaps most intriguingly, an ``allosteric'' algorithm has recently been implemented in disordered isostatic bond networks to alter their structure in order to maximize some fitness criterion~\cite{rocks2016designing,yan2016architecture}---polarization could serve as just such a criterion, raising the question of whether disordered lattices could achieve the same level of directionality as periodic Maxwell frames. They have already been shown to have edge modes with similar exponential decay~\cite{sussman2016topological,yan2016edge}.

Extending our linear analysis to nonlinear definitions is of clear interest. Initial examinations of two-dimensional systems suggest that the linear mode extends in a regular way to somewhat nonlinear deformations~\cite{rocks2016designing,rocklin2015transformable}. However, the nonlinear extension edge mode of the 1D rotor chain has been revealed to be a soliton~\cite{chen2014nonlinear,zhou2016kink}, meaning that finite bond swellings can generate and drive modes that spatially separate from the swelling itself. Conversely, origami chains~\cite{chen2016topological} are governed by the same linear map, Eq.~(\ref{eq:onedchain}), as the rotors but do not possess a soliton, demonstrating that linear analysis cannot predict nonlinear results conclusively.

In summary, Green's functions demonstrate the directionality of response in topologically polarized lattice and serve as  interesting means of examining other, related systems.

\section*{Acknowledgments}

The author is grateful to Bryan G. Chen, Michael Lawler, Xiaoming Mao and Jim Sethna
 for helpful conversations.
The author gratefully acknowledges  support from the the ICAM postdoctoral fellowship, the Bethe/KIC Fellowship, and the National Science Foundation Grant No. NSF DMR-1308089.

\section*{Appendix A: characterization of system via rigidity and compatibility matrices}

In this Appendix we describe how a linear mechanical system, and in particular a periodic one, can be described in terms of a set of linear maps.

We consider as our archetypal system a network of particles connected via Hookean springs. Consider two such particles initially located at $\pos_1, \pos_2$ and undergoing small, linear displacements $\uvec_1, \uvec_2$. Only the component of the displacement that projects along the direction pointing from one particle to the other, $\hat{\mathbf{r}}_{12}$ contributes to the spring extension:

\begin{align}
\ext_{12} = \hat{\mathbf{r}}_{12}\cdot \left(\uvec_2-\uvec_1\right).
\end{align}

\noindent We can combine all of our displacements into a single vector describing the total configuration of our system, $\uvec = \left(\uvec_1, \uvec_2,\ldots,\uvec_{\ndof}\right)$, and our displacements via $\evec = \left(\ext_1, \ext_2,\ldots \ext_{\ncon}\right)$. In this way, all of the above relationships may be expressed in terms of a single linear map, the rigidity (also ``compatibility'' or ``kinetic'') matrix:

\begin{align}
\evec = \rmat \uvec.
\end{align}

Similarly, we may determine the forces on particles from their spring tensions. For springs with unit spring constant, the tension is simply the (negative) extension and so the force on the first site, as defined above, is

\begin{align}
\fvec_1 = \ext_{12} \hat{\mathbf{r}}_{12}.
\end{align}

\noindent As before, we can combine these into a single map between bond extensions and forces on sites, known as the equilibrium matrix. However, because of the above relationships the equilibrium matrix is simply the transpose of the rigidity matrix:

\begin{align}
\fvec = \rmat^T \evec.
\end{align}

Combined, these give us the full dynamics. In terms of the dynamical matrix, $\dmat = \rmat^T \rmat$, the energy is

\begin{align}
E = \frac{1}{2} \evec^2 = \frac{1}{2} \left( \rmat \uvec \right)^2 =  \frac{1}{2} \uvec^T \dmat \uvec.
\end{align}

And the relationship between forces and displacements is simply

\begin{align}
\fvec = \dmat \uvec.
\end{align}

We now wish to introduce periodic structure, with a lattice subject to periodic boundary conditions running $\nsite_j$ sites along lattice vector $\latvec_j$ in the $j^\textrm{th}$ direction. As such, the $\alpha^\textrm{th}$ particle in the $\indreal^{\textrm{th}}$ cell is 
:

\begin{align}
\pos_{\indreal}^\alpha = \basisvec^\alpha + \sum_j \ireal_j \latvec_j.
\end{align}

Assuming periodic modes of the form $\uvec_\indreal = \uvec \exp \left( i \indrecip \cdot \indreal\right)$ with $\ireal_j = 0, 1, \ldots, \nsite_j-1$ and $\irecip_j = 0, 1, \ldots, \nsite_j-1$ means that our rigidity matrix above will become a function of wavenumber, since springs connect to particles in different cells. The wavenumber is flipped when relating forces to extensions, and $\rmat \rightarrow \rmat_\indrecip, \rmat^T \rightarrow \rmat^T_{-\indrecip}$.

The advantage of expressing the systems mechanical relations in this way (the same advantage present for many local, linear relationships) is that it diagonalizes our relations, such that $\uvec_\indrecip = \rmat_\indrecip \evec_\indrecip$ rather than $\uvec_\indrecip = 
\sum_{\indrecip'} \rmat_{\indrecip,\indrecip'} \evec_{\indrecip'}$. In this way, a large system is reduced to a large number of wavenumber-dependent small systems. Thus, we may invert the above relationships to determine the displacements resulting from a set of forces or of bond extensions, or the extensions resulting from forces:

\begin{align}
\uvec_\indreal &=& \frac{1}{\prod_j \ireal_j} \sum_{\indrecip_j \ge 0}^{\indrecip_j \le \nsite_j -1}
 e^{i \indrecip \cdot \left(\indreal-\indreal_s\right)}\rmat^{-1,g}_\indrecip \esource \\
\uvec_\indreal &=& \frac{1}{\prod_j \ireal_j} \sum_{\indrecip_j \ge 0}^{\indrecip_j \le \nsite_j -1}
 e^{i \indrecip \cdot \left(\indreal-\indreal_s\right)}\dmat^{-1,g}_\indrecip \fsource \\
\evec_\indreal &=& \frac{1}{\prod_j \ireal_j} \sum_{\indrecip_j \ge 0}^{\indrecip_j \le \nsite_j -1}
 e^{i \indrecip \cdot \left(\indreal-\indreal_s\right)}(\rmat_\indrecip^T)^{-1,g} \fsource
\end{align}

\noindent Here, the ``$-1,g$ '' indicates a Moore-Penrose generalized pseudo-inverse. It is easily calculated in terms of the singular value decomposition, even for non-square matrices. It gives either the actual solution to the original linear relationships or the solution that minimizes the ``error''. 

We now consider a large system, $\nsite_j \gg 1$. This does not, however, imply that the distance from source to target, $\indreal - \indreal_s$ is large. The relationships then become

\begin{align}
\uvec(\indreal) &=& \frac{1}{\left(2\pi\right)^\dim}\int_{-\infty}^{\infty} d^\dim \indrecip \,
 e^{i \indrecip \cdot \left(\indreal-\indreal_s\right)}\rmat^{-1,g}(\indrecip) \esource \\
\uvec(\indreal) &=& \frac{1}{\left(2\pi\right)^\dim}\int_{-\infty}^{\infty} d^\dim \indrecip \,
 e^{i \indrecip \cdot \left(\indreal-\indreal_s\right)}\dmat^{-1,g}(\indrecip) \fsource \\
\evec(\indreal) &=& \frac{1}{\left(2\pi\right)^\dim}\int_{-\infty}^{\infty} d^\dim \indrecip \,
 e^{i \indrecip \cdot \left(\indreal-\indreal_s\right)}(\rmat^T)^{-1,g}(\indrecip) \fsource
\end{align}

\section*{Appendix B: Topology, zero modes and response functions}

In this Appendix, we show how the direction of the response to a point source is determined by the topological polarization. To begin, we review how this polarization is related to the edge modes.

Consider a spring in crystal cell $\indreal$ connecting one of that cell's sites to a site in the cell whose first index is one higher. Based on our mode, this implies a spring extension of the form

\begin{align}
\ext_{12}(\indrecip) = \hat{\mathbf{r}}_{12} \left( e^{i \irecip_1} \uvec_2(\indrecip) - \uvec_1(\indrecip)\right).
\end{align}

In this way, we see that the rigidity matrix's elements are simply powers of $\{z_j\} \equiv \{e^{i \irecip_j}\}$. These complex numbers must have unit magnitude to satisfy periodicity or normalization. However, for finite lattices with open boundary conditions, missing or swollen bonds periodicity is broken and $|z_j| \ne 1$ modes become relevant. The condition for such zero modes is $\det(\rmat)=0$, leading to a polynomial in $\{z_j\}$ whose order is set by the number of bonds between unit cells in various lattice directions.

In this way, we may readily obtain the modes existing on edges with missing bonds parallel to the $\latvec_j$ lattice direction. We merely choose particular values of the remaining $z_{-j}$ (in particular, we can use all $ |z_{-j}| = 1$ modes as a basis of the lattice) and solve for the values of $z_j$ that satisfy $\det(\rmat)=0$. The fundamental theorem of algebra requires that we obtain one such mode (on one edge for  $|z_{j}| <1$, on the other for  $|z_{j}| >1$ or in the bulk for  $|z_{j}| = 1$) for each missing bond, ensuring that each edge mode permitted by the index theorem given in the main text.

Although obtaining the exact complex wavenumber requires solving the above algebraic equation, we can determine how many zero modes rest on which edge purely by considering the winding number of $\det(\rmat)=0$ across the bulk modes. The result is a topological polarization 

\begin{align}
\toppol_j = -\frac{1}{2 \pi i} \int_{\indrecip \rightarrow \indrecip + \recipvec_j}
\frac{\partial}{\partial \irecip_j} \log \textrm{det} \left[\rmat(\indrecip)\right],
\end{align}

\noindent which gives the number of zero modes on the $+\latvec_j$ edge (and the opposite on the  $-\latvec_j$ edge) in excess of that indicated by the local coordination number. This formulation follows from the Argument Principle of complex analysis, which in turn depends on the number of poles enclosed by the contour along $|z_{j}| = 1$. Those poles depend on how we choose to assign bonds to crystal cells (a gauge choice); our above result applies to a symmetrical choice and would otherwise need to be shifted.

Now, though, we can set aside all notion of an edge and simply consider evaluating the response functions of the previous appendix, which similarly require an integral around $-\pi < \irecip_j \le \pi$ or around the contour  $|z_{j}| = 1$. This is evaluated using the contour of the second figure of the main text. Ignoring the contributions at $\indrecip = 0$ (which are dealt with in detail in the main text), the response comes from the poles of $\rmat^{-1}(\indrecip)$, which are the zeroes of $\rmat(\indrecip)$ (the elements of the inverse of an $n\times n$ matrix are combinations of products of $n-1$ elements of the original matrix divided by its determinant).

Depending on which side of the source we wish to calculate the response, it is limited to zeroes at $|z_j|>(<)1$ (for $\ireal_j <(>) \ireal_j^s$). Hence, the bulk response depends on the same non-periodic zero modes as the edge modes and the topological polarization. In particular, a lattice that is fully polarized in the sense of, e.g., $|z_j|>1$ for all its modes, will have displacements responding to a swollen bond only on the $-\latvec_j$ side of the bond, growing until they reach the bond and then vanishing.

The analysis is similar for the compatibility matrix, $\rmat^T(-\indrecip)$, which has zeroes at $z_j \rightarrow z_j^{-1}$, revealing that in the above case spring extensions can only be induced on the $\latvec_j$ side of forces applied to sites. However, in this case displacements are generated as well, as indeed is the only way to induce spring extensions absent externally-imposed swelling. Similarly, the dynamical matrix $\dmat(\indrecip) = \rmat^T(-\indrecip) \rmat^T(\indrecip)$ has pairs of zero modes at $z_j$ and self stresses at $z_j^{-1}$, both of which correspond to states of mechanical equilibrium.

\section*{Appendix C: The one-dimensional chain}

In this Appendix, we present a more detailed derivation of the mechanical response of the 1D topological chain, whose rigidity matrix is given by the relation between extensions $\ext_\ireal$ and site modes $\disp_\ireal$

\begin{align}
\label{eq:onedchainapp}
e_\ireal = a \disp_\ireal - b \disp_{\ireal-1}.
\end{align}

\noindent More generally, this represents a one-dimensional lattice system with the most general uniform linear mapping from a mode $\uvec$ to a set of energetically costly deformations $\evec$. Requiring that no such deformations occur leads to a zero mode of the form

\begin{align}
\label{eq:onededgemode}
\disp_\ireal = \left(\frac{b}{a}\right)^{\ireal-1}\disp_s.
\end{align}

\noindent 
For a system such as a simple ball and spring chain, which we term ``critical'', $a = b$ and there is a uniform zero mode, in that case a mode of translation. More generally, though, this is an edge mode, exponentially decaying as one moves left (right) for $b/a>1$ ($b/a<1$). Such an edge mode is possible in a finite system consisting of $\length$ modes and $\length-1$ interactions between them. We, however, consider systems subject to periodic boundary conditions, in which no such zero modes exist and the above equation is invertible. The result, as may be verified by direct algebraic evaluation, is

\begin{align}
\disp_\ireal = \frac{1}{a^\length-b^\length} \sum_{\ireal'=1}^{\length} a^{\textrm{mod}(\ireal' - \ireal -1,\length)}
b^{\textrm{mod}(\ireal - \ireal',\length)} \ext_{\ireal'}.\quad
\end{align}

This complete inversion, which is impossible for the general case in which zero modes yet exist, allows us to construct the explicit response of the system to a localized source $\evec = \ext_s \delta_{\ireal',\ireal_s}$. Such a source may be thought of as changing the length of a particular spring, either by altering its equilibrium length or by applying a force. The resultant displacement is

\begin{align}
\label{eq:onedmode}
\disp_\ireal=
\frac{1}{a} \left( \frac{b}{a} \right)^{\ireal - \ireal_s}
\begin{cases}
      \frac{1}{1- \left(b/a\right)^\length}
			& \ireal \ge \ireal_s \\
			\frac{1}{\left(a/b\right)^\length-1}
			& \ireal < \ireal_s
   \end{cases}.
\end{align}

Although this direct method of solution is perfectly satisfactory when it exists, it does not illustrate the topological nature of the system, which lies in the winding of its phonon bands. To see this, we work in reciprocal space, where our rigidity matrix is simply $\rmat(\irecip) = a  - b \exp(-2 \pi i \irecip)$. Using the orthonormal basis of reciprocal space, the mode resultant from the local bond swelling is

\begin{align}
\disp_\ireal =
 \frac{1}{\length} \sum_\irecip 
\frac{
\exp \left[
2 \pi i \irecip \left(\ireal - \ireal_s\right)/\length
\right]
}{a  - b \exp(-2 \pi i \irecip/\length)}.
\end{align}

\noindent In the limit of large system sizes, this becomes

\begin{align}
\disp_\ireal = \frac{1}{2 \pi} \int_{-\infty}^{\infty} d \irecip \, \frac{\exp\left[i \irecip \left(\ireal -\ireal_s\right)\right]}
{a  - b\exp(-i \irecip)}.
\end{align}

This is the simplest possible form for our contour integration technology, leading to a mode

\begin{align}
\disp_\ireal =
\frac{1}{a} \left(\frac{b}{a}\right)^{\ireal - \ireal_s} \textrm{sign}(\ireal - \ireal_s) \Theta \left[
\left(\ireal - \ireal_s \right) \left(a-b\right)\right].
\end{align}

\bibliography{gfuncbib}

\end{document}